\documentclass [10pt]{iopart}
\usepackage[dvips]{graphicx}

\begin{document}
\jl{4}

\title[Effect of Ad-Hoc Vehicular Network on Traffic Flow]
{Effect of Ad-Hoc Vehicular Network on Traffic Flow: Simulations in the Context of Three-Phase Traffic Theory}

\author{Boris S. Kerner $^1$, Sergey L. Klenov $^2$, A. Brakemeier $^1$}      


\address{
$^1$ Daimler Research, GR/PTI,
HPC: 050 - G021, D-71059 Sindelfingen, Germany 
}

\address{
$^2$ Moscow Institute of Physics and Technology, Department of Physics, 141700 Dolgoprudny,
Moscow Region, Russia
}

\date{October 02, 2009}


\begin{abstract}
 Effects of vehicle-to-vehicle (or/and vehicle-to-infrastructure communication,
 called also V2X communication)  on traffic flow, which are relevant for ITS, are numerically studied. To make   the study
    adequate with real measured traffic data, a testbed for wireless vehicle communication based on a microscopic model in the framework 
 of three-phase traffic theory is developed and discussed. In this testbed, vehicle motion in traffic flow 
 and analyses of a vehicle communication channel access based on IEEE 802.11 mechanisms, 
 radio propagation modeling, message reception characteristics as well as all other effects associated
  with ad-hoc networks are integrated into a three-phase traffic flow model. Thus simulations of both 
  vehicle ad-hoc network and traffic flow are integrated onto a single testbed and perform simultaneously. 
  This allows us to make simulations of ad-hoc network performance as well as diverse scenarios of the 
  effect of wireless vehicle communications on traffic flow during simulation times, which can be 
  comparable with real characteristic times in traffic flow. In addition, the testbed allows us to 
  simulate cooperative vehicle motion together with various traffic phenomena, like traffic 
  breakdown at bottlenecks, moving jam emergence, and a possible effect of danger warning massages about the breakdown vehicle
  on traffic flow.
\end{abstract}

\maketitle

\section{Introduction}

Wireless vehicle communication, which is the basic technology for ad-hoc vehicle networks, 
is one of the most important scientific fields of future ITS. This is because there are many 
possible applications of ad-hoc vehicle networks, including various systems for danger warning, 
traffic adaptive assistance systems, traffic information and prediction in vehicles, improving of 
traffic flow characteristics through adaptive traffic control, etc.~\cite{Ad-hocA9,Ad-hocB9,Ad-hocC9,Ad-hocD9,Ad-hocE9,Ad-hocF9,Ad-hocG9}. 
However, the evaluation of ad-hoc vehicle networks requires many communicating vehicles moving in 
real traffic flow, i.e., field studies of ad-hoc vehicle networks are very complex and expensive. For this reason, to 
prove the performance of ad-hoc vehicle networks based on wireless vehicle communication, reliable simulations of ad-hoc 
vehicle networks are of great importance and indispensable.

\begin{figure}[!t]
\centering
\includegraphics[width=2.5in]{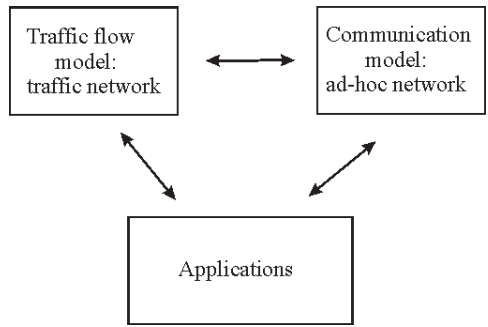}
\caption[]{Usual scheme of testbeds for simulations of ad-hoc vehicle networks
 (e.g.,~\cite{Schmitz9,Torrent19,Torrent29,Schmidt19,Choffnes9,Maurer9,Chen9,Workshop20079,Sklar9}).}
\label{Scheme_usual}
\end{figure}

An usual schema for the development of a testbed for simulations of ad-hoc vehicle networks includes a traffic flow model, 
a model for vehicle communications that is often based on the use of ns-2 simulator~\cite{ns-2}, and application models 
(applications in Fig.~\ref{Scheme_usual}) (e.g.,~\cite{Schmitz9,Torrent19,Torrent29,Schmidt19,Choffnes9,Maurer9,Chen9,Workshop20079,Sklar9}). 
Application models determine, for example, necessary changes in vehicle 
behavior in traffic flow after receiving of the associated message or/and whether this message should be resent to other vehicles or not. 
There are two different networks in such testbeds: (i) a traffic network simulated with the use of the traffic flow model and 
(ii) a communication (ad-hoc) network simulated with the use of the communication model in which positions and other 
characteristics of each communicated vehicle are taken from simulations of the traffic network made at the latest 
point in time. Simulations of many communicating vehicles in the communication network with known communication 
models are very time intensive. For this reason, often the model of communication network (communication model in Fig.~\ref{Scheme_usual}) 
performs simulations based on traffic flow data previously simulated through the use of the traffic flow model
(off-line simulations of traffic networks).  In some of these testbeds, to study applications in which vehicle 
behavior should be changed in accordance with received messages, the communication model performs simulations 
after each time step of traffic flow simulations. In any case, the use of this simulation schema (Fig.~\ref{Scheme_usual})
 requires a very long run time of the simulations, which can be some order of magnitude longer than real time of vehicle moving in traffic flow.
 
 In this article, we perform a numerical study of 
 possible effects of vehicle-to-vehicle or/and vehicle-to-infrastructure communication
 (called also V2X communication)  on traffic flow. To make   the study
    adaquate with real measured traffic data, a testbed for wireless vehicle communication based on a microscopic model in the framework 
 of three-phase traffic theory is developed and discussed.
 In this testbed, simulations of a 
 traffic network and an ad-hoc vehicle network as well as applications are integrated into 
 an united network, i.e., there is only one network in this testbed. The network describes 
 both vehicle motion in traffic flow and communications as well as the effect of applications 
 on traffic flow and vehicle behavior. As a result, simulations of ad-hoc performance and various 
 applications can be made many times quicker than with the scheme shown in Fig.~\ref{Scheme_usual}. To reach this goal, 
 each vehicle in this network exhibit different attributes needed for both vehicle motion and communications, 
 and application scenarios.   In addition, we should note that recently based on a study of measured data on 
 many highways in different countries a three-phase traffic theory has been developed. In contrast with earlier 
 traffic flow theories and models, three-phase traffic theory can explain and predict all known empirical features 
 of traffic breakdown and resulting congested patterns~\cite{KernerBook,KernerBook_2}. For this reason, 
 we use a traffic flow model in the framework of three-phase traffic theory.
A  model for vehicle ad-hoc network is presented in Sect. III. Simulations of three scenarios of C2C application devoted to ad-hoc network 
  influence   on traffic flow are presented in Sects.  IV--VI, respectively. However,
  in section $\lq\lq$Backgroughs" we briefly consider  channel access mechanism used for simulations of vehicle communication
  (Sect.~\ref{Channel}) and some features of three-phase traffic theory used for traffic simulations (Sect.~\ref{3Phase}).
  
 \section{Backgrounds} 
 
 \subsection{Channel Access Mechanism used for Simulations of Vehicle Communication \label{Channel}}
 
  If there are messages to be sent and the medium is free, the vehicle sends the message 
 that has the highest priority and/or is the first one in the message queue in this vehicle.
To prevent  collisions between messages sent by
different communicating vehicles, a message access method
is usually applied. 

\begin{figure}[!t]
\centering
\includegraphics[width=3.3in]{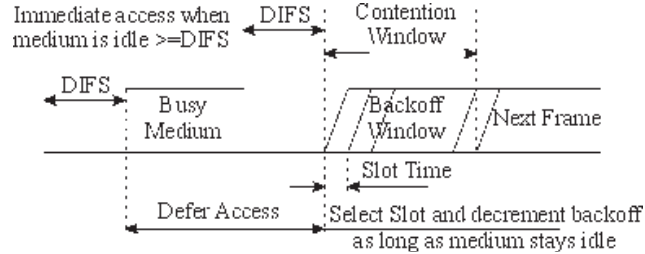}
\caption[]{IEEE 802.11 basic access mechanism~\cite{IEEE802_1_9,IEEE802_2_9,IEEE802_3_9}  
}
\label{ad-hoc2}
\end{figure}

As an example, we use here   IEEE 802.11e basic access method~\cite{IEEE802_1_9,IEEE802_2_9,IEEE802_3_9}.
No access is possible when medium is  busy.
After the medium has been free, in accordance with the IEEE 802.11e access method, there is a
backoff procedure applied  for each of the communicating vehicles independently of  each other.
At the end of the backoff procedure, a decision whether
 the medium is free or busy is made.
 
 In accordance with IEEE 801.11e mechanism, we can summarize       possible cases as follows:

 (i)      None of the signal powers in the matrix is greater than a signal
     receiving threshold (RXTh); 
then no message is   accepted. 
 Under this condition, there can be two possible cases:

 (a)  
 The sum of all signal powers in the matrix is smaller than the carrier sense 
 threshold CSTh. 
 Then the medium is free; therefore, the abovementioned
 backoff procedure is applied for the message sending (Fig.~\ref{ad-hoc2}).

 (b) 
  The sum of all signal powers in the matrix is equal to or greater than the threshold 
 CSTh. Then the medium is busy for the vehicle.
 
  (ii)  The greatest signal 
 power of the signal powers in the matrix is greater than the threshold 
 RXTh. Under this condition, it is tested
for the matrix of signal powers whether 
the ratio between the power of this greatest 
signal power and the sum of the powers of all other signals stored in the matrix is greater than the required 
signal-to-noise ratio (SNR) 
at the selected data rate (DR) for the whole duration of the 
message:
 1)  
If yes, then the signal could be considered to be received. 
2) 
Otherwise, there is no message acceptance at this time instance.
 
 \subsection{Three-Phase Traffic Theory -- The Basis for Update Rules of Vehicle Motion
 \label{3Phase}}
 
 There are three phases in three-phase traffic theory~\cite{KernerBook}:
\begin{itemize}
\item	Free flow (F).
\item	Synchronized flow (S).
\item	Wide moving jam (J).
\end{itemize}
The synchronized flow and wide moving jam traffic phases are associated with congested traffic.

The empirical study of real measured traffic data mentioned above shows that  
there are common pattern features that are qualitatively the same independent of 
 highway infrastructure, weather, 
 percentage of long vehicles,   vehicle technology, etc.
 The  empirical traffic phase definitions [S] and [J] for the synchronized flow and wide moving jam phases 
 in congested traffic made in three-phase traffic theory are these common empirical pattern features.
 
 The definition of the wide moving jam phase [J]: A wide moving jam is a moving jam that maintains the mean velocity 
of the downstream jam front, even when the jam propagates through other traffic phases or bottlenecks. 

The definition of the synchronized flow phase [S]: In contrast with the wide moving jam phase, 
the downstream front of the synchronized flow phase does not exhibit the wide moving jam characteristic feature; 
in particular, the downstream front of the synchronized flow phase is often fixed at a bottleneck. 

Recall that a moving jam is a propagating upstream localized structure of great vehicle density and very low speed 
spatially limited by two jam fronts. Within the downstream jam front vehicles accelerate escaping from the jam; 
within the upstream jam front, vehicles slow down approaching the jam. 

  \begin{figure}[!t]
\centering
\includegraphics[width=3.5in]{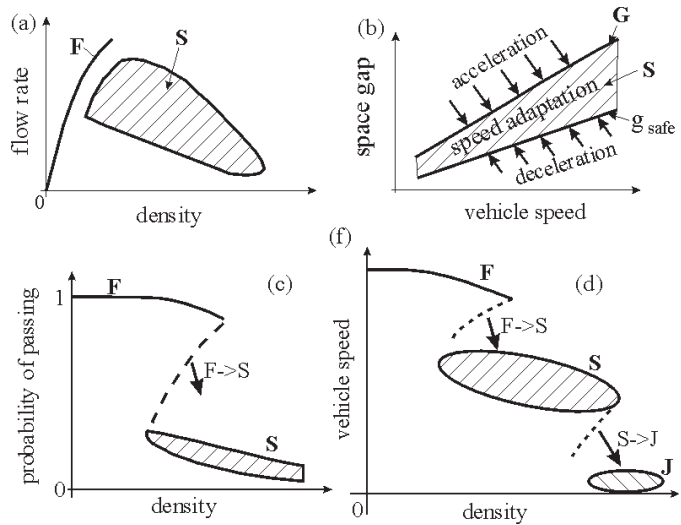}
\caption[]{Hypotheses of three-phase traffic theory about traffic breakdown and
wide moving jam emergence~\cite{KernerBook}.}
 \label{Hyp} 
\end{figure}

 \begin{figure}[!t]
\centering
\includegraphics[width=3.5in]{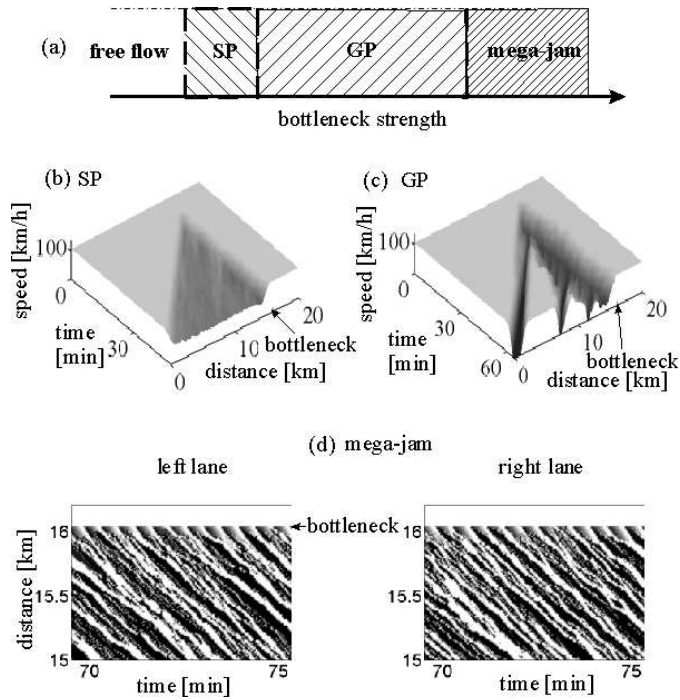}
\caption[]{Hypotheses of three-phase traffic theory about congested traffic patterns resulting from
traffic breakdown: (a) Simplified diagram of congested patterns. (b--d)
SP (b), GP (c), and mega-jam (d). In (d), within black regions vehicles are in standstill, whereas within white regions vehicles
move at very low speeds within the mega-jam~\cite{KernerBook,KernerBook_2}.}
 \label{Hyp2} 
\end{figure}
 
 The phase definitions [S] and [J] in real measured traffic data are the empirical basis of hypotheses of three-phase traffic theory
  inmlemented into mathematical three-phase traffic flow models.
 Some of these hypotheses are as follows:
 
 (1) Rather than a fundamental diagram, in three-phase traffic theory
steady states\footnote{Steady states of synchronized flow
are hypothetical states in which vehicles move at the same time-independent speed and with the same space gaps
to each other.} of synchronized flow   
 cover a two-dimensional (2D) region in the flow--density plane (dashed region in 
 Fig.~\ref{Hyp} (a)):   While adapting speed to the speed of the preceding vehicle,
 a driver accepts different space gaps  within a limited gap range $g_{\rm safe}\leq g\leq G$, where
 $G$ and $g_{\rm safe}$ are some synchronization and safe space gaps, respectively 
(Fig.~\ref{Hyp} (b))). 
 
(2) Traffic breakdown at a highway bottleneck is a local phase transition from free flow to synchronized flow
(F$\rightarrow$S transition). Traffic breakdown is explained by a Z-shaped density function of
probability of vehicle passing  (Fig.~\ref{Hyp} (c)): There is a drop in this probability when
free flow transforms into synchronized flow.

(3)
In free flow,
wide moving jams emerge    due to a sequence of F$\rightarrow$S$\rightarrow$J
 transitions that can be illustrated by 
   a double Z-characteristic  (Fig.~\ref{Hyp}  (d)).
 The first  Z-shaped relationship, which includes
    free flow  $F$ and  synchronized flow $S$, is associated with an F$\rightarrow$S transition, i.e., traffic breakdown 
     (labeled by arrows F$\rightarrow$S in Figs.~\ref{Hyp} (c, d)).
   The second Z-shaped relationship, which includes
   synchronized flow $S$ and low speed states within  wide moving jams $J$,
    is associated with an S$\rightarrow$J transition (labeled by arrow S$\rightarrow$J in Fig.~\ref{Hyp} (d)).

(4)
 If the   bottleneck strength, which characterizes
  the influence of a bottleneck on traffic breakdown and
  resulting traffic congestion, 
  increases gradually, 
  firstly a synchronized flow pattern 
  (SP) emerges   upstream of the bottleneck
  (Fig.~\ref{Hyp2} (a, b)). Congested traffic within an SP 
  consists of   synchronized flow only. 
  At a greater bottleneck strength, the SP transforms 
   into a general congested pattern
(GP) (Fig.~\ref{Hyp2} (a, c)).  Congested traffic within an GP consists of synchronized flow and wide moving jams that emerge  in 
synchronized flow.
If the  bottleneck strength increases further, 
 the GP transforms into a  mega-wide moving jam (mega-jam)
 (Fig.~\ref{Hyp2} (a, d)).

\section{United Ad-Hoc Network Model}

\begin{figure}[!t]
\centering
\includegraphics[width=2.5in]{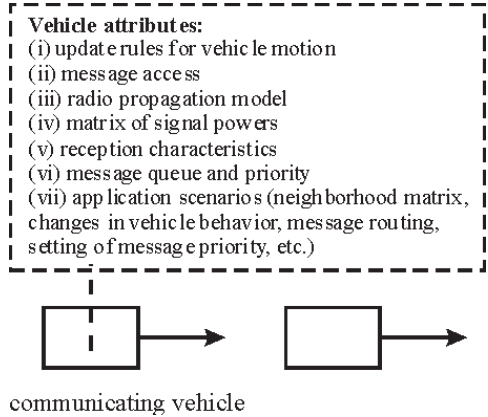}
\caption[]{Scheme of  simulations of ad-hoc networks and traffic flow within a united network model.}
 \label{ad-hoc1} 
\end{figure}

In the united network model of traffic flow, C2C-comunications, and ad-hoc networks, 
there are dynamic vehicle attributes, which exhibit each of the communicating vehicles (Fig.~\ref{ad-hoc1}). 
All other vehicles in the network, which cannot communicate, exhibit only one dynamic attribute: 
update rules for vehicle motion. If in addition with communicating vehicles the network includes 
roadside communication units (RSU), each RSU exhibits the communicating vehicle attributes with 
the exception of the update rules for vehicle motion. 

\subsection{Update Rules for Vehicle Motion \label{Motion}}

The vehicle attribute $\lq\lq$update rules for vehicle motion" 
are given by a stochastic microscopic three-phase traffic flow model of Kerner and Klenov~\cite{KKl,KKl2003A}, 
which as shown in~\cite{KernerBook,KernerBook_2}
 can explain all fundamental measured spatiotemporal features of real traffic flow. Because the detailed model description can be found in~\cite{KernerBook},
basic rules of vehicle motion in the model are in Appendix~\ref{Ap}.
 
 \subsection{Message Access \label{access}}
 
 During a motion of a communicating vehicle 
in a traffic network, the vehicle (and RSU)  attribute $\lq\lq$message access"
 calculates vehicle access possibility 
for message sending for each vehicle independent of each other in an unsynchronous manner, i.e., in contrast with update rules of vehicle motion
no fixed time discretization is used in the  model of vehicle communication.
In the version of the testbed, we use IEEE 802.11e basic access method of Sect.~\ref{Channel}.

 \subsection{Radio Propagation Model} 
 
 Based on the vehicle (and RSU) attribute $\lq\lq$radio propagation model", 
 signal powers of the message that has been sent by the vehicle are calculated 
 for current locations of all other communicating vehicles and RSUs. 
  
  There are many different radio propagation models. However, at a time instant the real  
       signal power of the message   sent by a vehicle at a location of another communicating vehicle
       can be an extremely complex function of urban infrastructure (e.g., whether there are   buildings causing
    strong  signal reflection effects), the current vehicle distribution on the road (e.g., how
     many vehicles are between the vehicle and  the location as well as
    whether there are long-vehicles between the vehicle and   location at which 
    the signal power should be found), etc. 
    
    One of the approaches 
    to solve this complex problem   is as follows. In our model, each communicating vehicle 
     can apply either one of many  radio propagation models stored in the vehicle or one of the different parameters of a radio propagation model.
    At a given time instant, the choice of the radio propagation model or the model parameter occurs automatically
     for each vehicle individually and independently of radio propagation models used by other vehicles. This radio propagation model choice
     is based on the current vehicles' distribution on the road (and if known, urban infrastructure). Because a set
     of radio propagation models and their variable parameters stored in vehicles should cover diverse scenarios of different urban infrastructures
     and vehicle distributions, the radio propagation models   should be associated with {\it field study measurements}
     made in accordance with these possible different scenarios. Unfortunately, at this time there is no such  detailed experimental basis for
     the development of   the model set available. 
     
     For this reason, as long as the abovementioned experimental basis is not available,
     in simulations
     we use one of the simplest radio propagation models --
      a well-known two-ray-ground radio propagation model  
  \begin{equation}
  P(R)= P_{0}(R_{0}/R)^{m}, 
  \label{radion}
  \end{equation}
where $R_{0}$ is a communication range, $m$ is a model parameter that is $m\geq 2$,
$R$ is the distance between two communicating vehicles, $P$ is the signal power, $P_{0}$ is constant.

In (\ref{radion}), the communication range $R_{0}$ and value $m$ are in general
time-dependent model parameters. This is because $R_{0}$ and $m$ 
 depend on the current vehicles' distribution on the road 
as well as on urban characteristics (e.g., buildings and other obstacles in the neighberhood of the road)
causing the reflection, diffraction, and other effects that influence on the signal power. 
Because the vehicle distribution on the road and some urban characteristics can randomly  
change over time during vehicle motion, 
the values $R_{0}$ and $m$ in (\ref{radion}) can be stochastic time variables.
Time-dependencies of $R_{0}$ and $m$ in (\ref{radion}) 
for different vehicles can be very different. Nevertheless, due to vehicle motion we can expect
that {\it mean} characteristics of the stochastic variables of $R_{0}$ and $m$ in (\ref{radion}) can be 
the same for different communicating vehicles.

\subsection{Matrix of Signal Powers \label{Matrix}} 

 In the model  (Fig.~\ref{ad-hoc1}), to make the decision whether
 the medium is free or busy or else the vehicle has received a message or not,
 based on the vehicle  attribute $\lq\lq$radio propagation model"~\cite{KKB1_9}, 
 signal powers of   messages   sent by  all other communicating vehicles
 are calculated. 
 If a signal power is greater than a given threshold signal power denoted by $P_{\rm th}$   (model parameter), 
 then this signal power of 
 the associated message is stored into a $\lq\lq$matrix of signal powers" of the   vehicle:
   \begin{itemize}
\item 
 at each time instant, the matrix  of signal powers of the vehicle
contains signal powers
 of messages sent by other vehicles
in ad-hoc network
that are greater than the threshold signal power $P_{\rm th}$ at the location of the vehicle.
 \end{itemize}
 This threshold  $P_{\rm th}$ is chosen to be  much smaller than a carrier sense threshold  
  (CSTh). The smaller  $P_{\rm th}$ is chosen, the greater the accuracy of simulations 
 of ad-hoc network performance, however, the longer the simulations run time. 
Signal reception characteristics 
(whether the medium is free or busy  as well as   whether the vehicle    has received a message or not)
are associated with an analysis of the     
 matrix of signal powers, which is 
automatically made at each time instant for each communicating vehicle   individually.
The decision about signal collisions is further used for a study of ad-hoc network performance. 

\begin{table}[!t]
\renewcommand{\arraystretch}{1.3}  
\caption{Hypothetical example for matrix of signal powers}
\label{table}
\centering
\begin{tabular}{|c||c||c||c||c||c|}
\hline
ID of sending & & & & & \\
 vehicle    &   25  & 382  & 37 & 36 & 31   \\ 
\hline
Distance (in [m]) & & & & &\\
 between & & & & &\\
 the receiving &  234  & 345   & 300  & 70  & 562    \\
vehicle 33 and & & & & &\\
 sending vehicle & & & & &\\
\hline
Received signal & & & & &\\ 
power (in [dBm]) & & & & &\\ 
of the message & $-$ 91    & $-$ 95     & $-$ 93     & $-$ 81     & $-$ 99       \\ 
sent   at the location  & & & & &\\
 of the vehicle 33 & & & & &\\
\hline
\end{tabular}
\end{table}

We consider an application of the matrix of signal powers for
   a hypothetical 
   example\footnote{In this example as well as in simulations presented below,
   we have used the model (\ref{radion}) with the communication range 
  $R_{0}=$ 200 m, $P_{0}=10^{-9}$ mW, RXTh $=-$ 90 dBm, 
  CSTh $=-$ 96 dBm, SNR $=-$ 6 dBm.} of
 a communicating
 vehicle with a vehicle-ID (identification number) 33 (Table~\ref{table}).
 In this case, in the matrix   of the vehicle 33 there are several signal powers
of those messages sent
 at the time $t$ by other vehicles  in an ad-hoc network
  whose signal powers are   greater than the threshold 
 CSTh $=-$ 96 dBm
  at the location of the vehicle  33.
  However, only the  signal power of a message sent by vehicle ID 36 that
  is equal to  $-$ 81 dBm     is   greater than
  a signal receiving threshold 
 RXTh $=-$ 90 dBm.
 The ratio between the power of this greatest 
signal power of a message sent by vehicle ID 36 
  and the sum of the powers of all other signals stored in the matrix is greater than the required 
signal-to-noise ratio (SNR)  
 for the whole duration of the message.
Thus in  
 the matrix of signal powers  
 the signal sent by  vehicle 36 that is 70 m outside of the location of vehicle 33
could be considered to be received by vehicle 33\footnote{In  simulations presented below,  we have used
  $P_{\rm th}=-$ 116 dBm, 
 which allows us to have a good balance between accuracy and  
 simulation time. Simulation results are changed in the range of 
 about 1$\%$, when instead of  $P_{\rm th}=-$ 116 dBm, the threshold $P_{\rm th}=-$ 126 dBm has been used.}.

\subsection{Reception Characteristics}

Signal reception characteristics are associated with an analysis of the  matrix of signal powers of Sect.~\ref{Matrix}, which is 
automatically made at each time instant for each communicating vehicle   individually.
In particular, this   matrix   
is used for the decision whether the medium is free or busy at each time instant as well as for the decision whether the vehicle    has received a message or not.

We see that at each time instant the matrix of signal powers 
is used both for the decision whether the vehicle  has received a message and whether there are collisions 
between two or more different signals at the current vehicle location. Message collisions 
are realized for example, if there are two or 
more signals within the matrix and the highest power is greater than the threshold RXTh, 
however, based on the above procedure the decision has been made that there 
is no message acceptance at the time instance. 
The decision about signal collisions is further used for a study of ad-hoc network performance.

\subsection{Message Queue and Priority}

Based on an application, which should be simulated,
 in the model each communicating vehicle (or RSU) exhibits an attribute
  of message queue organization and individual message priority performance governed
   automatically. Because each communicating vehicle or RSU manages these features individually, 
   this attribute can be chosen differently for various types of the communicating vehicles or RSUs.

\subsection{Application Scenarios}

In the model, each communicating vehicle (and RSU) exhibits an attribute $\lq\lq$application scenario".  
This attribute governs the organization of all messages that are received and to be sent. Based on this 
attribute and the message context just received by the vehicle, the vehicle can change its behavior 
in traffic flow (e.g., the vehicle slows down or changes the lane, or else changes the route, etc.).

 \section{Effect of Danger Warning   $\lq\lq$Breakdown Vehicle Ahead"
on Congested Patterns}

\begin{figure}[!t]
\centering
\includegraphics[width=3.3in]{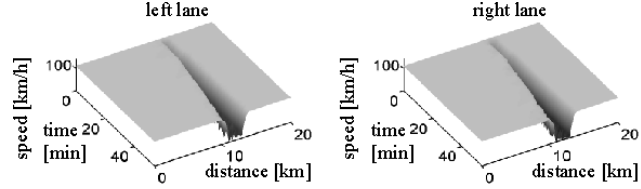}
\caption[]{Simulation of traffic congestion under breakdown vehicle ahead
  without communication. }
 \label{BreVeh_NO} 
\end{figure}

We consider an application scenario in which due to the  $\lq\lq$breakdown" of one of the vehicles, this vehicle
has to decelerate and comes to a stop in the right lane at location 12.5 km of a two-lane road.
After a driver moving initially in the right lane recognizes the breakdown vehicle, it changes to the left lane.
We assume that the   distance at which vehicles see this breakdown vehicle and therefore begin to change lane is equal to 100 m.
Simulation results of the average vehicle speed (left figure for the left lane  and right figure for the right lane)
are presented in Fig.~\ref{BreVeh_NO} for the flow rate in an initial free flow upstream $q_{\rm in}=$ 1125 vehicles/h/lane.
We see that when there is {\it no} communicating vehicles, traffic congestion occurs caused by the breakdown vehicle ahead.

\begin{figure}[!t]
\centering
\includegraphics[width=3.3in]{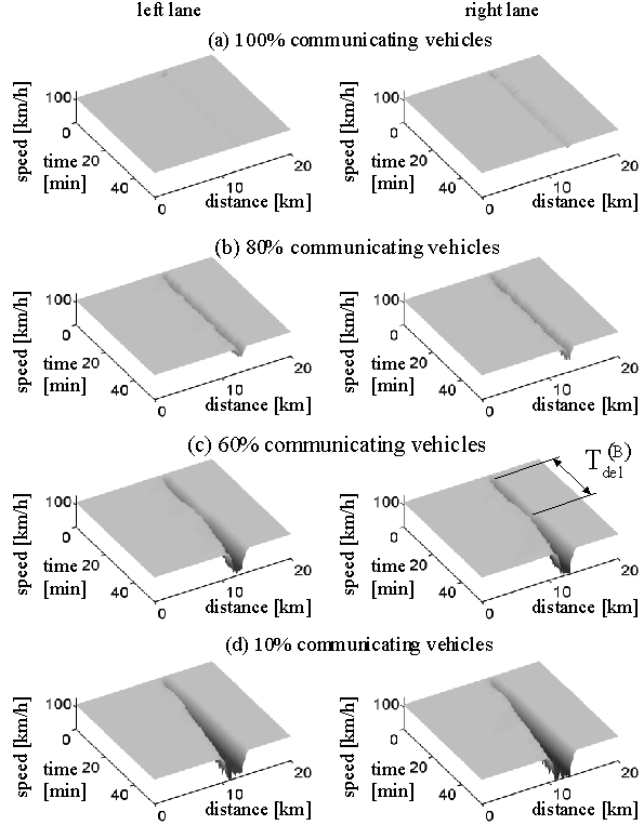}
\caption[]{Simulation of effect on traffic congestion of danger warning for different
pecentage of communicating vehicles }
 \label{BreVeh_COM} 
\end{figure}

Now we consider the same scenario for the case of communicating vehicles, which sent $\lq\lq$dange warning" message
about the breakdown vehicle ahead. We assume that after the communicating vehicles have received this message they
increase the  distance to 600 m at which the communicating
vehicles moving in the right lane  try to change to the left lane.
Because we  assume that vehicles, which cannot communicate, begin to  change the lane at the   distance 100 m, the 
simulation results depend on the percentage of the communicating vehicles as presented in Fig.~\ref{BreVeh_COM}
\footnote{The 
breakdown vehicle is not shown in Figs.~\ref{BreVeh_NO}
and~\ref{BreVeh_COM}.}.

We can see that there is a critical percentage of the communicating vehicles that is about 70$\%$ (Fig.~\ref{BreVeh_COM}):
	(i) if the   percentage of the communicating vehicles is greater than the critical one, then no
	traffic congestion occurs. (ii) 
	Otherwise, a congested pattern occurs whose downstream front is fixed at the location of the breakdown vehicle; characteristics of this congested patterns
	are similar to those as for the case when no communication vehicles moving in traffic flow (Fig.~\ref{BreVeh_NO}).
	However, we should note that when the   percentage of the communicating vehicles  is smaller but close to the critical value, then
	there is a random time delay in the occurrence of the congested pattern that is denoted by $T_{\rm del}^{\rm (B)}$ in Fig.~\ref{BreVeh_NO} (c).
	The time delay is a random value: at the same simulation parameters but different initial spatial distributions of traffic variables
	  very different values $T_{\rm del}^{\rm (B)}$ are found.

 \section{Prevention of Traffic Breakdown at On-Ramp Bottleneck Through Vehicle
Ad-Hoc Network \label{car_car_traffic} }

In accordance with three-phase traffic theory~\cite{KernerBook,KernerBook_2}, we can assume that there can be  the following 
two hypothetical   possibilities to prevent traffic breakdown
at an on-ramp bottleneck   through changes in driver behavior of communicating vehicles:

(i) A decrease  in
the amplitude of disturbances on the main road
occurring when vehicles merge from on-ramp onto the right lane of the main road.
This decreases the probability of  nucleus occurrence   required for traffic breakdown.
 
 (ii) An increase in probability of over-acceleration.

 \begin{figure}[!t]
\centering
\includegraphics[width=3.3in]{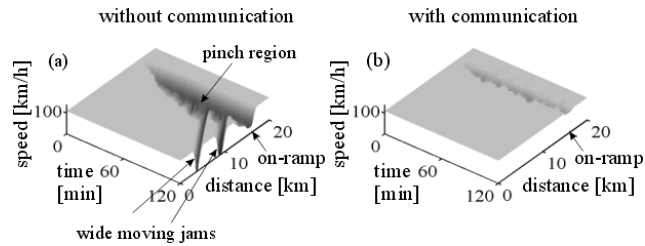}
\caption[]{Simulations of prevention of traffic breakdown at on-ramp bottleneck through
vehicle communication:  Speed in time and space without communication (a)
 and with vehicle communication (b). 
 \label{ad-hoc5} }
\end{figure}

In simulations (Fig.~\ref{ad-hoc5}),
 there is an on-ramp bottleneck at location 16 km. 
The flow rate on the main road upstream of the bottleneck is  $q_{\rm in}=$
1827 vehicles/h/lane; the flow rate to the on-ramp is $q_{\rm on}=$ 600 vehicles/h. 
At these flow rates  if there is no C2C communication, a general congested pattern
 (GP)~\cite{KernerBook} occurs at the bottleneck
  (Fig.~\ref{ad-hoc5} (a)). The GP consists of a pinch region of synchronized flow (labeled by $\lq\lq$pinch region")
and     wide moving jams upstream of the pinch region (labeled by $\lq\lq$wide moving jams").
 
  Now we assume (Fig.~\ref{ad-hoc5} (b)) that   all vehicles are communicated vehicles, which try to send a non-priority message with time intervals 0.1 s.   Vehicles moving
 in the on-ramp lane send  a  priority message for neighbor vehicles moving in the right road lane
when the vehicle intends to merge  from the on-ramp onto the main 
road.
We assume that the following vehicle in the right lane   
  increases a time headway for the vehicle merging. Simulations show that in comparison with the case in which
  no vehicle communication is applied and the GP occurs (Fig.~\ref{ad-hoc5} (a))
  this change in driver behavior of communicating vehicles decreases   disturbances in free flow at the bottleneck. This results in the prevention of traffic breakdown
  (Fig.~\ref{ad-hoc5} (b)).

\section{Effect of Ad-Hoc Vehicle Network
on Congested Traffic Patterns}

\begin{figure}[!t]
\centering
\includegraphics[width=3.3in]{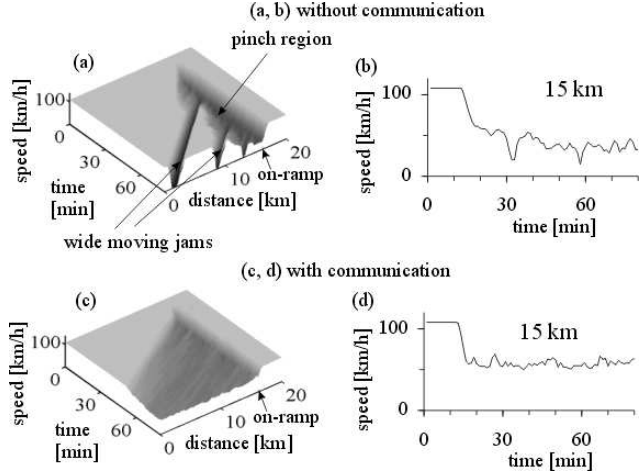}
\caption[]{Simulation of the effect of vehicle
 communication on congested patterns:   Speed in time and space (a, c) and   at a  location $x=$ 15 km 
 (b, d) without communication (a, b)
 and with  communication (c, d). 
 \label{ad-hoc4} }
\end{figure}

Here we consider a case of the same communicating vehicles as that in Sect.~\ref{car_car_traffic} at
    $q_{\rm in}=$
1946 vehicles/h/lane when   traffic control
through the use of changes in driver behavior  in free flow at the bottleneck discussed above  is not applied. In this case,
  traffic breakdown occurs at the bottleneck resulting in
 GP occurrence (Fig.~\ref{ad-hoc4} (a, b)). 
  In accordance with three-phase traffic theory~\cite{KernerBook,KernerBook_2}, we can assume that there can be
    the following two hypothetical possibilities to prevent moving jam emergence in synchronized flow
  through    changes in driver behavior of communicating vehicles moving in synchronized flow:

(i) A decrease  in
the amplitude of disturbances in synchronized flow upstream of the bottleneck.
This decreases the probability of  nucleus occurrence   required for the emergence of wide moving jams.

(ii) A decrease in the density of synchronized flow  upstream of the bottleneck. This decreases
the critical speed required for the emergence of wide moving jams in synchronized flow.
The lower the critical speed, the smaller the
probability for the emergence of wide moving jams.

  We assume that
  after  synchronized flow has just occurred due to traffic breakdown
  at the bottleneck, communicating vehicles, which reach the synchronized flow,
  send  priority messages about the speed reduction to vehicles moving in free flow upstream.
Each message comprises a minimum space gap  that should be maintained by vehicles while 
  moving in the synchronized flow. 
 
In vehicle motion rules of the model, the associated change in driver behavior is simulated through 
an increase in probability $p_{1}$  in (\ref{prob1}) from  0.3 for vehicles, 
which have no information about the required space gap to  0.55 for the  vehicles that received the message. The greater $p_{1}$,
  the greater the difference between vehicle space gap and 
  a safe space gap and, therefore,   the less the probability for moving jam emergence in the synchronized flow~\cite{KernerBook}. 
As a result of space gap increase within the synchronized flow, 
 at the same flow rates upstream of the bottleneck as those in Fig.~\ref{ad-hoc4} (a, b)   
rather than the GP a widening synchronized flow pattern (WSP) is forming (Fig.~\ref{ad-hoc4} (c, d)).  
Whereas in the pinch region of the GP the mean space gap is 15 m, 
it is 25 m within the WSP. Due to the transformation of the GP 
into the WSP, two effects are achieved: (i) wide moving jams do not occur and
 (ii) the average speed within synchronized flow upstream of the bottleneck 
 increases from about 40 km/h within the GP to 60 km/h within the WSP. 
 These effects can result in a considerable increase in the efficiency and safety of traffic.

\section{Conclusion}

1. Simulations made with the use of the testbed for ad-hoc networks presented 
in this paper allow us	to perform quick simulations of various applications of C2C-communication and 
ad-hoc network performance associated with the real behavior of vehicular traffic.  This is due of the following advantages of this testbed: 
 As in a real ad-hoc network, there is only one network in the testbed in which 
C2C-communication, ad-hoc performance, and traffic flow characteristics are simulated
 simultaneously during vehicle motion. This testbed feature decreases the simulation run
  time considerably and exhibits a sufficient accuracy of simulations. 
This testbed feature allows us to make an easier understanding of ad-hoc 
network and traffic flow performances associated with those applications in which message 
contexts should influence on vehicle behavior. This is crucial especially for communication 
based safety systems that currently are studied in various research projects
 (e.g. WILLWARN~\cite{Ad-hocE9} and SAFESPOT~\cite{SAFESPOT}).

2. Simulations show that  changes in driver behavior made through the use of
ad-hoc vehicle network can indeed prevent traffic breakdown and/or lead to the dissolution of moving jams.
Thus C2X communication can increase the efficiency and safety of traffic considerably.

\appendix
\section{Stochastic Three-Phase Traffic Flow Model \label{Ap}}

\begin{figure}[!t]
\centering
\includegraphics[width=2.3in]{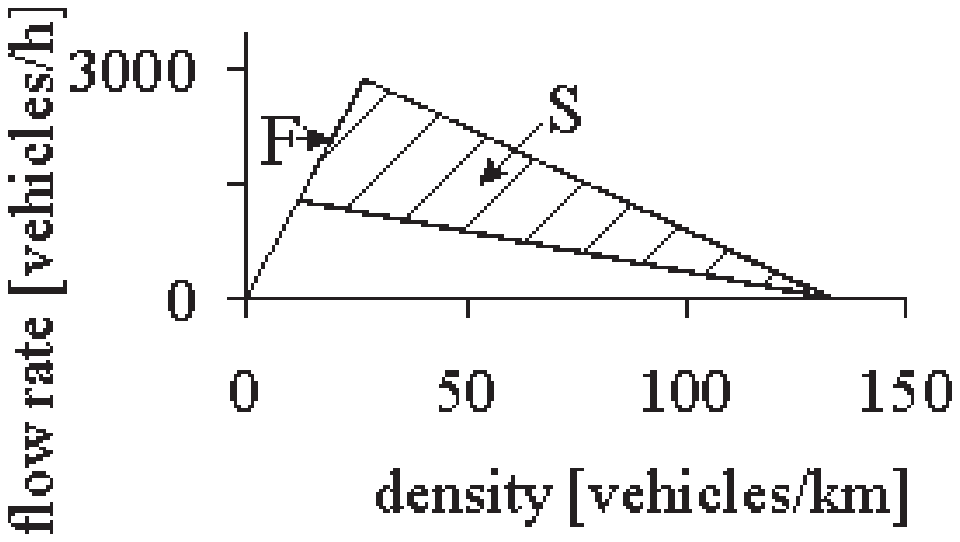}
\caption[]{Steady speed states for the Kerner-Klenov stochastic three-phase
traffic flow in the   flow--density plane 
}
\label{KKl_Steady}
\end{figure}

Basic rules of vehicle motion in the model are as follows~\cite{KKl,KKl2003A}:
\begin{equation}
v_{n+1}=\max(0, \min({v_{\rm free}, \tilde v_{n+1}+\xi_{n}, v_{n}+a \tau, v_{{\rm s},n} })),
\label{final}
\end{equation}
\begin{equation}
\label{next_x}
x_{n+1}= x_{n}+v_{n+1}\tau,
\end{equation}
\begin{eqnarray}
\label{next}
\tilde v_{n+1}=\max(0, \min({v_{\rm free}, v_{{\rm c},n}, v_{{\rm safe},n} })), \\
 \\
v_{c,n}=\left\{
\begin{array}{ll}
v_{n}+\Delta_{n} &  \textrm{at $g_{n}
\leq G_{n}$} \\
v_{n}+a_{n}\tau &  \textrm{at $g_{n}> G_{n}$},
\end{array} \right.
\label{next1}
\end{eqnarray}
where
\begin{equation}
\Delta_{n}=\max(-b_{n}\tau, \min(a_{n}\tau, \ v_{\ell,n}-v_{n})),
\label{next2}
\end{equation}  
$v_{\rm free}$ is the maximum speed 
in free flow that is  constant,   $g_{n}=x_{\ell, n}-x_{n}-d$ is the space gap, $x_{n}$ is the vehicle co-ordinate, the lower index $\ell$
marks functions and values related to the preceding vehicle;   
all vehicles have the same length $d$;
index $n$ corresponds to the discrete time $t=n\tau$,
$n=0,1,2,...;$ $\tau$ is time step;
a synchronization space gap $G_{n}$ and
a safe speed $v_{\rm safe,n}$ have been   discussed in~\cite{KernerBook} in detail.
Steady states of this model, i.e., states in which all vehicles move at a time-independent speed at
the same space gap between each other cover a 2D-region in the flow--density plane (Fig.~\ref{KKl_Steady}).

Random deceleration and acceleration $\xi_{n}$ in (\ref{final}) are applied depending on 
whether the vehicle decelerates or accelerates, or else maintains its speed:
\begin{equation}
\xi_{n}=\left\{
\begin{array}{ll}
-\xi_{\rm b} &  \textrm{if $S_{n+1}=-1$} \\
\xi_{\rm a} &  \textrm{if  $S_{n+1}=1$} \\
0 &  \textrm{if  $S_{n+1}=0$},
\end{array} \right.
\label{noise}
\end{equation}
where  
$S$ in (\ref{noise})  denotes the state of motion ($S_{n+1}=-1$ represents
   deceleration, $S_{n+1}=1$    acceleration, and $S_{n+1}=0$  
motion at nearly constant speed)
\begin{equation}
S_{n+1}=\left\{
\begin{array}{ll}
-1 &  \textrm{if $\tilde v_{n+1}< v_{n}-\delta$} \\
1 &  \textrm{if $\tilde v_{n+1}> v_{n}+\delta$} \\
0 &  \textrm{otherwise},
\end{array} \right.
\label{noise_}
\end{equation}
where  
$\delta$ is  constant 
($\delta \ll a\tau$).
 \begin{equation} 
 \xi_{\rm a}=a\tau \Theta (p_{\rm a}-r),
 \label{xi_acc} 
 \end{equation}
 where $p_{\rm a}$ is probability of random acceleration, $a$ is the maximum acceleration, $r={\rm rand (0,1)}$,
 $\Theta (z) =0$ at $z<0$ and $\Theta (z) =1$ at $z\geq 0$; 
\begin{equation} 
 \xi_{\rm b}=a\tau \Theta (p_{\rm b}-r),
 \label{xi_dec} 
 \end{equation}
Random acceleration $a_{n}$ and deceleration 
$b_{n}$ are  
 \begin{equation} 
 a_{n}=a \Theta (P_{\rm 0}-r_{1}),
 \label{r_acc} 
 \end{equation} 
 \begin{equation} 
 b_{n}=a \Theta (P_{\rm 1}-r_{1}),
 \label{r_dec} 
 \end{equation}
where
 the probabilities $P_{0}$ and $P_{1}$
are 
\begin{equation}
P_{0}=\left\{
\begin{array}{ll}
p_{0}(v_{n}) & \textrm{if $S_{n} \neq 1$} \\
1 &  \textrm{if $S_{n}= 1$},
\end{array} \right.
\label{prob0}
\end{equation}
\begin{equation}
P_{1}=\left\{
\begin{array}{ll}
p_{1} & \textrm{if $S_{n}\neq -1$} \\
p_{2}(v_{n}) &  \textrm{if $S_{n}= -1$},
\end{array} \right.
\label{prob1}
\end{equation}
$r_{1}={\rm rand (0,1)}$, speed functions for probabilities $p_{0}(v_{n})$ and $p_{2}(v_{n})$ are considered
in~\cite{KernerBook}; $p_{1}$ is constant.
Lane changing rules, models of highway bottlenecks and other model parameters in all simulations presented in the article
 are listed in Table~16.11 of the book~\cite{KernerBook}.

  \section*{Acknowledgments}

We thank Gerhard N{\"o}cker, Andreas Hiller and Christian Weiss for fruitful discussions.





\end{document}